Chromatin Structure Changes in Human Disease: A Mini Review

Yuriy Shckorbatov

V.N. Karazin Kharkiv National University, Kharkiv, 61022, Ukraine
yuriy.shckorbatov@gmail.com

Abstract:

There are many experimental data indicating the correlations of the changes in high level of organization of chromatin in human cells and changes in the state of the whole organism related to disease, state of tiredness or aging. In our previous work: arXiv.org-2018 (1812.00186) we analyzed the publications on the topic up to 2017. In this work we focused on works upon the problem of connection of the state of chromatin with human diseases published in 2018-2019. In the modern literature the most attention is paid to problem of chromatin transformations in different forms of cancer, Alzheimer's disease, and hereditary diseases. Summing up, the tendency of scientific research of noncommunicable diseases is shifting towards investigation of aspects of nuclear regulation of disease origin, connected with conformation of chromatin.

Introduction

Supramolecular organization of chromatin in human cells is connected with the state of cell and consequently with the state of the whole organism. Since the previous article on the topic: "Chromatin in human cells as characteristics of the state of human organism" reviewing the literature mainly up to 2016 (1), some works concerning changes in chromatin organization at different diseases have appeared. These works are concentrated around the theme "Chromatin structure changes in human disease". Here we present a short review of research works on the topic in 2017-2019.

Discussion

Several reviews on different aspects of chromatin structure modifications connected with human diseases appeared in 2018-2019. General view on alteration of nuclear structure connected with hereditary syndromes and cancer one can see in (2). The separate section in this review deals with chromatin transformations in disease.

The role of chromatin rearrangements in the development of nervous system development with a special focus on trithorax related proteins. The role of different

ATP-dependent chromatin remodeling complexes in in different syndromes, for instance, schizophrenia, microcephaly, intellectual disability, and Rett-like phenotypes, is also analyzed (3). Also, the precise investigation of the role of ISWI chromatin remodeling complexes is done in connection with brain development disorders (4).

The review of experimental works investigating molecular events at chromatin level connected with Alzheimer's disease (AD), Huntington's disease (HD) and amyotrophic lateral sclerosis (ALS) is presented in (5). The age-related changes are also investigated. The attention of authors of review is focused on histone H3 modifications: disease-related acetylation and methylation of histone H3.

Aspects of histone deacetylases (HDACs) possible usage in medical treatment of fibrosis–connected diseases are discussed. Deacetylation of histones in chromatin, as known, modulates transcriptional activity regulating chromatin compaction. Certain types of HDACs are involved in origin of various kinds of fibrosis. Using results obtained from animal models authors discuss the possible application of HDAC inhibitors in medical treatment of human diseases associated with fibrosis (6).

The role of topoisomerases in chromatin remodeling in disease and normal aging is discussed in review (7). The authors focus on two topoisomerases: Top2α, operating with double-stranded DNA, and Top3β, changing topology of DNA and RNA. Both enzymes are required for normal heterochromatin formation and function. The individuals bearing mutations of Top3β are characterized by the shortened lifespan and neurological disorders. Authors analyze the facts of heterochromatin loss in AD, and loss of HP1 or H3K9 methylation in heterochromatin connected to cancer progression and tumorigenesis (7).

The role of deficiency in enzymes modulating chromatin in such diseases as retinoblastoma, diabetic retinopathy, glaucoma is demonstrated in (8).

Chromatin changes at molecular level in human brain associated with AD and aging are discussed in (9). An epigenome-wide association study using the histone 3 lysine 9 acetylation (H3K9ac) mark which marks transcriptionally active open chromatin, was done in 669 samples of human prefrontal cortices. Participants of this investigation agreed to donate their brain upon death. As accumulation of tau and amyloid-β proteins are connected to AD. The acetylation of histone H3 in cortices in connection with these protein was performed. In contrast with amyloid-β, tau protein accumulation had a broad effect on the epigenome, affecting histone 3 lysine 9 acetylation. A tau-induced alterations of the histone 3 lysine 9 acetylation were much more profound than the changes that are attributable to amyloid pathology. The effect of these changes influence the transcriptome. Thus, this work presents data on precise localization and distribution of chromatin sites characterized by presence of histone 3 acetylated form - H3K9ac. This may be an informative approach in research of chromatin state in aging and disease.

The informative review on topologically associating domains (TAD) and also CCCTC-binding factor (CTCF) binding cites changes in cancer is presented in (10). The experimental works demonstrating disruptions of TAD boundaries in case of neuroblastoma, medulloblastoma, and leukemia are analyzed. Proposed models are discussed, among them: deletion of boundaries resulting in unusual promoter-enhancer contacts and inversions in TADs alter the contents of neighboring TADs. The boundary integrity can be influenced by hyper-methylation of CpG and GC-rich CTCF binding motif. This resulted in reducing of CTCF binding in glioma cells. Therefore, the data analyzed in this review support the notion of important role of TAD changes in cancer.

In the review paper (11) changes in TAD, connected to limb formation disorders and cancer are discussed. In authors view, disrupting TAD boundaries can ectopically activate neighboring proto-oncogenes which induce carcinogenesis.

Several experimental works deal with the role of condensin protein complexes in the etiology of diseases is investigated. Chromosome compaction in mitosis is essential for their segregation during mitosis. In vertebrates, two condensin complexes ensure timely chromosome condensation, sister chromatid disentanglement, and maintenance of mitotic chromosome structure (12). The mice bearing mutation in gene *Ncaph2* coding subunit H2 of condensin II complex, are hypomorphic, have reduced brain size, mitoses are characterized by frequent anaphase chromosome bridge formation. Frequent chromosome bridges also observed in mitosis of condensin-deficient cells derived from patient with microcephaly bearing mutations in condensin I and II subunits genes *NCAPD2, NCAPH,* and *NCAPD3*. Chromosome bridges formation results in micronucleus formation and aneuploidy in daughter cells (12).

Chromosomal instability is a hallmark of cancer, but mitotic regulators are rarely mutated in tumors. Mutations in the condensin complexes, which restructure chromosomes to facilitate segregation during mitosis, are significantly enriched in cancer genomes, but experimental evidence implicating condensin dysfunction in tumorigenesis is lacking. The role of condensin II complex subunit mutation in T-cell lymphoma is elucidated in (13). Mice with mutations in a condensin II subunit (*Caph2$^{nes}$*) leads to T-cell lymphoma. Ploidy in such mice was severely perturbed at the $CD4^+CD8^+$ T-cell stage. T cells show violations during chromosome segregation, resulting DNA damage and elevated ploidy in daughter cells. Tumors bearing *Caph2* mutation are near diploid but carry deletions spanning tumor suppressor genes. *P53* inactivation results in proliferation of *Caph2* mutant cell lines with aneuploidy and structural rearrangements in chromosomes resulting in highly aggressive disease.

A new possible way of chromatin compaction regulation in disease is connected with histone glycation (14). In model experiments authors show that glycation by methylglyoxal (MGO) at lower concentrations induces relaxation of chromatin fiber, while higher MGO concentrations induce their compaction. Authors identify glycation of histones in breast cancer cells, xenografts and tumors, but not in normal cultured cells. In authors opinion the obtained results suggest an additional pathogenesis-connected mechanism of cellular metabolic damage (14).

Conclusion

In recent time (2018-2019) the problem of chromatin activity regulation via its conformational transitions in the development of various diseases attracts attention of many researchers. The most attention is paid to different forms of cancer, Alzheimer's disease, and hereditary diseases. The most popular methods of investigation are connected with genomics and determining of expression of enzymes related to chromatin modification, but the methods of TAD investigation in disease are also proliferating. Thus, the epigenetic aspects of disease origin, connected with conformation of chromatin on supramolecular level are of increasing importance.

References

(1) Shckorbatov Y. Properties of Chromatin in Human Cells as Characteristics of the State of Human Organism: A Review. arXiv preprint arXiv:1812.00186. 2018 Dec 1.
(2) Northcott JM, Weaver VM. Altered Nucleus and Disease. In: Nuclear Architecture and Dynamics 2018 Jan 1 (pp. 493-512). Academic Press.
(3) Moccia A, Martin DM. Nervous system development and disease: a focus on trithorax related proteins and chromatin remodelers. Molecular and Cellular Neuroscience. 2018 Mar 1;87:46-54.
(4) Goodwin LR, Picketts DJ. The role of ISWI chromatin remodeling complexes in brain development and neurodevelopmental disorders. Molecular and Cellular Neuroscience. 2018 Mar 1;87:55-64.
(5) Berson A, Nativio R, Berger SL, Bonini NM. Epigenetic regulation in neurodegenerative diseases. Trends in neurosciences. 2018 Sep 1;41(9):587-98.
(6) Yoon S, Kang G, Eom GH. HDAC inhibitors: therapeutic potential in fibrosis-associated human diseases. International journal of molecular sciences. 2019 Jan;20(6):1329.
(7) Lee SK, Wang W. Roles of Topoisomerases in Heterochromatin, Aging, and Diseases. Genes. 2019 Nov;10(11):884.
(8) Popova EY, Barnstable CJ. Insights Into the Epigenetics of Retinal Development and Diseases. InEpigenetics and Regeneration 2019 Jan 1 (pp. 355-383). Academic Press.
(9) Klein HU, McCabe C, Gjoneska E, Sullivan SE, Kaskow BJ, Tang A, Smith RV, Xu J, Pfenning AR, Bernstein BE, Meissner A. Epigenome-wide study uncovers large-scale changes in histone acetylation driven by tau pathology in aging and Alzheimer's human brains. Nature neuroscience. 2019 Jan;22(1):37.

(10) Kaiser VB, Semple CA. When TADs go bad: chromatin structure and nuclear organisation in human disease. F1000Research. 2017;6.

(11) Bruneau BG, Nora EP. Chromatin Domains Go on Repeat in Disease. Cell. 2018 Sep 20;175(1):38-40.

(12) Martin CA, Murray JE, Carroll P, Leitch A, Mackenzie KJ, Halachev M, Fetit AE, Keith C, Bicknell LS, Fluteau A, Gautier P. Mutations in genes encoding condensin complex proteins cause microcephaly through decatenation failure at mitosis. Genes & development. 2016 Oct 1;30(19):2158-72.

(13) Woodward J, Taylor GC, Soares DC, Boyle S, Sie D, Chathoth K, Vukovic M, Tarrats N, Jamieson D, Campbell KJ, Blyth K. Condensin II mutation causes T-cell lymphoma through tissue-specific genome instability. Genes & development. 2016 Oct 1;30(19):2173-86.

(14) Zheng Q, Omans ND, Leicher R, Osunsade A, Agustinus AS, Finkin-Groner E, D'Ambrosio H, Liu B, Chandarlapaty S, Liu S, David Y. Reversible histone glycation is associated with disease-related changes in chromatin architecture. Nature communications. 2019 Mar 20;10(1):1289.